\documentclass[conference]{IEEEtran}

\usepackage{cite}
\usepackage{amsmath,amssymb,amsfonts}
\usepackage{algorithmic}
\usepackage{graphicx}
\usepackage{textcomp}
\usepackage{xcolor}
\usepackage[hyphens]{url}

\def\BibTeX{{\rm B\kern-.05em{\sc i\kern-.025em b}\kern-.08em
    T\kern-.1667em\lower.7ex\hbox{E}\kern-.125emX}}

\usepackage[labelfont=bf]{caption}
\usepackage{subcaption}
\usepackage{hyperref}
\usepackage{comment}

\newcommand{\ignore}[1]{}



\newcommand{\name}{SCuBA}       
\newcommand{\fullname}{\underline{S}tatic \underline{Cu}da \underline{B}ounds \underline{A}nalyzer}
\newcommand{\tool}{{\name{}}}                          

\newcommand{\cuda}{CUDA}

\usepackage[frozencache,cachedir=.]{minted}
\setminted{linenos=true,autogobble,xleftmargin=15pt}
\newenvironment{code}
{\minted[fontfamily=cmtt,escapeinside=@@,fontsize=\scriptsize]{cuda}}
{\endminted}

\newenvironment{pycode}
{\minted[fontfamily=cmtt,escapeinside=@@,fontsize=\scriptsize]{python}}
{\endminted}

\newenvironment{assemcode}
{\minted[fontfamily=cmtt,escapeinside=@@,fontsize=\scriptsize]{gas}}
{\endminted}

\usepackage{tikz}
\newcommand{\circled}[1]{\tikz[baseline=(char.base)]{
            \node[shape=circle,draw,inner sep=0.5pt] (char) {#1};}}
\newcommand{\mycircled}[1]{\circled{\textbf{#1}}}
\newcommand{\myparagraph}[1]{\noindent\textbf{#1:}}

\newcommand{\ie}{i.e.,}
\newcommand{\eg}{e.g.,}

\newcommand{\aka}{a.k.a.,}

\newenvironment{mybullet}{\begin{list}{$\bullet$}
{\setlength{\topsep}{0mm}\setlength{\itemsep}{0mm}
\setlength{\parsep}{0mm}
\setlength{\listparindent}{\parindent} 
\setlength{\itemindent}{0mm}\setlength{\partopsep}{0mm}
\setlength{\labelwidth}{-2mm}
\setlength{\leftmargin}{0mm}}}{\end{list}}

\newcommand{\mysubsection}[1]{\subsection{#1}}

\newcommand{\textcode}[1]{{\small \texttt{#1}}}

\usepackage{multirow}

\newcommand{\cucatch}{cuCatch}
\newcommand{\csan}{Compute sanitizer}
\newcommand{\csanshort}{CSan}
\newcommand{\lmi}{LMI}
\newcommand{\gpuarmor}{GPUArmor}

\begin{document}

\pdfpagewidth=8.5in
\pdfpageheight=11in

\newcommand{\iscasubmissionnumber}{847}

\pagenumbering{arabic}

\title{Chasing Elusive Memory Bugs in GPU Programs}





\makeatletter
\newcommand{\linebreakand}{%
  \end{@IEEEauthorhalign}
  \hfill\mbox{}\par
  \mbox{}\hfill\begin{@IEEEauthorhalign}
}
\makeatother

\author{
  \IEEEauthorblockN{
    Anubhab Ghosh*,
    Ajay Nayak*,
    Dhananjay Rao Thallikar Shyam,
    Arkaprava Basu
  }
  \IEEEauthorblockA{\textit{Indian Institute of Science}}
}




\maketitle
\thispagestyle{plain}
\pagestyle{plain}


\begin{abstract}

Memory safety bugs, such as out-of-bound accesses (OOB) in GPU programs, can compromise the security and reliability of GPU-accelerated software.
We report the existence of \textit{input-dependent} OOBs in the wild that manifest \textit{only} under specific inputs.
All existing tools to detect OOBs in GPU programs rely on runtime techniques that require an OOB to manifest for detection.
Thus, input-dependent OOBs elude them. 
We also discover \textit{intra-allocation} OOBs that arise in the presence of logical partitioning of a memory allocation into multiple data structures. 
Existing techniques are oblivious to the possibility of such OOBs. 

We make a key observation that the presence (or absence) of semantic relations among program variables, which determines the size of allocations (CPU code) and those calculating offsets into memory allocations (GPU code), helps identify the absence (or presence) of OOBs. 
We build \tool{}, a first-of-its-kind compile-time technique that analyzes CPU and GPU code to capture such semantic relations (if present). 
It uses a SAT solver to check if an OOB access is possible under any input, given the captured relations expressed as constraints. 
It further analyzes GPU code to track logical partitioning of memory allocations for detecting intra-allocation OOB. 
Compared to NVIDIA's \csan{} that misses 45 elusive memory bugs across 20 programs, \tool{} misses none with no false alarms. 

\end{abstract}



\renewcommand{\thefootnote}{\fnsymbol{footnote}}
\footnotetext[1]{\vspace{-1em}Both authors contributed equally.}
\renewcommand{\thefootnote}{\arabic{footnote}}

\section{Introduction}
\label{sec:introduction}

A recent article by Google's Project Zero highlights memory safety bugs as a key vulnerability of the software ecosystem~\cite{google-safety}. 
It reports that over $40\%$ of the reported vulnerabilities are due to out-of-bound (OOB) accesses, even after decades of research in detecting them~\cite{valgrind,softbound,asan,RSan,cbmc,esbmc,sparc,cheri,cherivoke,cherix86,mpx,nofat,aos}.
An OOB occurs when a program accesses a memory allocation outside its bounds.
These bugs are common in programs written in memory-unsafe languages that allow pointer manipulations, such as C and C++.  
The OOBs often serve as doors to security vulnerabilities, and they can cause crashes and incorrect results, threatening software reliability \cite{google-safety,rop}.  

Today, GPU programs are a key component of the software ecosystem -- from AI/ML to HPC~\cite{google-cloud-gpu,aws-gpu}. 
The primary languages used to program GPUs, \eg{} \cuda{}~\cite{CUDAProgrammingGuide}, OpenCL~\cite{opencl}, HIP~\cite{amdhip}, and OpenACC~\cite{openacc}, are dialects of C++. 
Like C/C++, GPU programs written in these languages are vulnerable to memory safety bugs.
Recent works demonstrated that attackers can exploit OOBs in GPU programs to critically threaten the security and privacy of software~\cite{gpuexploitation,gpu-overflow,mindcontrolattack}. 

Several recent works focus on detecting memory safety bugs in GPU programs~\cite{computesanitizer, cucatch, gpuarmor, gmod, gpushield, letmein}. 
At a high level, they all follow the same basic principle -- maintain metadata for every memory allocation and instrument memory instructions in the kernel to detect safety violations at runtime. 
Some adopt a software-only approach~\cite{computesanitizer,cucatch,gmod,clarmor} while others propose new hardware to limit the overheads of checks at runtime~\cite{gpushield,gpuarmor,letmein}. 

\noindent\textbf{Limitations of prior works:} 
We notice that all existing techniques rely on runtime dynamic analysis.
Consequently, they suffer from a fundamental shortcoming -- they require \textit{memory safety bugs to manifest during execution} for detection. 
However, a bug may \textit{not} manifest in every execution of a buggy program. 
Arguably, the \textit{most critical bugs to catch are the elusive ones} -- those that do not manifest in many executions and thus, are more likely to escape software testing.
However, when they do manifest, these bugs (\eg{} OOBs) can cause unpredictable and intermittent disruptions. 

We demonstrate that such elusive OOBs exist in the wild! 
We discover the existence of tens of \textit{input-dependent} OOBs in open-source \cuda{} programs (kernels) and popular libraries. 
These OOBs manifest \textit{only} when a GPU-accelerated program is executed with specific inputs, \ie{} depends on the input.  

Unfortunately, existing techniques are highly susceptible to missing such elusive OOBs.
They would need a stroke of luck in the form of `right' inputs to manifest these bugs during testing, so that they can detect these input-dependent OOBs before the bugs may cause intermittent failures and/or open security vulnerabilities during production deployment.  

Next, we discover the possibility of a new type of bug in \cuda{} programs -- \textit{intra-allocation} OOB. 
Many \cuda{} programs logically partition a single buffer (memory allocation) into multiple different data structures. 
Such a program structure is often forced upon by the constraints of the programming language. 
For example, \cuda{} allows programmers to allocate \textit{only one} buffer in the scratchpad (shared memory) \textit{at runtime}, \aka{} dynamic shared memory~\cite{cudaShm}. 
If the program logic requires multiple data structures in the shared memory, programmers must logically partition the buffer (allocation).
In such programs, an intra-allocation OOB occurs when an instruction accesses a data structure other than the one it was supposed to, where both the data structures are carved out of the same memory allocation. 

The existing techniques, however, track only memory allocation boundaries and are agnostic of the partitioning of an allocation into multiple data structures. 
Consequently, they are incapable of detecting intra-allocation OOBs. 

Furthermore, prior software-only works~\cite{cucatch, computesanitizer, gmod} add significant performance and memory overhead due to the need to track allocation and memory accesses at runtime. 
Others propose significant new hardware to limit overheads~\cite{letmein, gpuarmor, gpushield}.
Both pose significant challenges to wider adoption. 

\noindent\textbf{Goals and Insights:} 
We set out to create a first-of-its-kind static analysis (compile-time) technique to detect elusive input-dependent and intra-allocation OOBs, without requiring hardware modifications and runtime overheads. 

The key challenge is to detect OOBs by analyzing only the source code during compilation when the actual inputs to the program are unknown. 
To this end, we make a key observation: bug-free CUDA programs naturally relate program variables that determine the sizes of memory allocations to the offsets (indices) of GPU memory instructions that access those allocations. 
An \textit{absence} of a semantic relation between these variables indicates the possibility of input-dependent OOBs. 


Importantly, the semantic relations are inferable from the source thanks to the use of common program variables. 
For example, it is typical for the host (CPU) code to decide memory allocation size based on input problem size (\eg{} matrix size).
Usually, the input size also determines the number of threads (\aka{} thread grid dimension) that the GPU program (\ie{} kernel) is launched with.
Each GPU thread would then use its unique thread identifier within the grid to derive the offset(s) into an allocation that it must compute on.  
The range of values of thread identifiers is bounded by the dimension of the thread grid. 
In effect, this binds the offsets of memory instructions to the size of the memory allocations it accesses. 
An absence of such relations among the variables indicates the presence of OOBs, even if it is yet to manifest in any execution of the program. 

\noindent\textbf{Software tool:} 
We use these observations to create a first-of-its-kind compile-time software tool, named \tool{} (\fullname{}), to detect elusive OOBs in \cuda{} programs. 
It analyzes the host code to infer the calculation for the size of memory allocations (\eg{} \textcode{cudaMalloc}). 
It analyzes kernel code to infer how each memory access instruction calculates the offsets into the memory allocations. 
\tool{} then creates a set of constraints from the relations among the relevant variables. 
\tool{} then invokes a SAT solver~\cite{sat-wiki} to determine if offsets of memory access can \textit{ever possibly} fall outside the corresponding allocation size, in the presence of the constraints.
If yes, an OOB is possible. 
Otherwise, not. 


Notice that \tool{} does \textit{not} rely on program inputs. 
Its analysis solely rests on the semantic relations among the program variables that determine memory allocation sizes and the offsets into them. 
Consequently, it detects input-dependent OOBs that may not have manifested yet in any execution. 

To detect intra-allocation OOBs, \tool{} analyzes the semantically-rich intermediate representation (IR) of \cuda{} programs.
It tracks pointer arithmetic operations that logically partition a single allocation into multiple data structures. 
It then uses the SAT solver to determine if the offset of a memory access can go past the logical boundary of the data structure it accesses under any possible input.

We implement \tool{} atop the MLIR framework~\cite{mlir}. 
It uses CGeist~\cite{polygeist} to compile \cuda{} programs (host and kernel) to MLIR's IR and uses the SAT solver from Google's Operations Research (OR) Tools~\cite{ortools}.
\tool{} found $30$ previously unreported and elusive input-dependent OOBs.
Overall, across a diverse set of $20$ workloads, \tool{} successfully detects $45$ OOBs, including intra-allocation OOBs, that NVIDIA's \csan{}~\cite{computesanitizer} fails to report.

\noindent\textbf{Contributions:} We make the following key contributions:
\mycircled{1} We demonstrate the existence of input-dependent OOBs in open-source \cuda{} programs and libraries that elude existing runtime techniques.
\mycircled{2} We discover the possibility of intra-allocation OOBs. 
\mycircled{3} We create a first-of-its-kind static analysis tool to accurately detect OOBs without runtime overheads. 

\section{Background}
\label{sec:background}

A streaming multiprocessor (SM) is the basic execution block on GPUs, consisting of multiple lanes that execute instructions on different data, and a scratchpad (a software-managed hardware cache). 
GPU has a hierarchy of caches and onboard memory.
GPU's onboard memory typically consists of High Bandwidth Memory (HBM) or GDDR technologies. 

GPUs are commonly programmed using dialects of C, and C++ such as \cuda{}~\cite{CUDAProgrammingGuide} or OpenCL~\cite{opencl}. 
We will focus on \cuda{}, without any loss of generality. 

\myparagraph{Structure of GPU-accelerated programs}
It consists of the host code (CPU) and the kernel that executes on the GPU. 
The host code orchestrates the overall execution of the program, which includes launching the kernel and allocating memory. 

The kernel consists of instructions that \textit{each} GPU thread executes. 
The thread is the smallest unit of execution in a kernel that occupies $one$ execution lane. 
A group of threads (\eg{} $32$) forms a warp that executes the same instruction on multiple execution lanes. 
A collection of warps forms a threadblock that resides on the \textit{same} SM. 
A grid consists of all the threadblocks that execute a GPU kernel. 

The host code allocates memory for data structures and prepares arguments for the kernel before launching it. 
The host also configures the number of threads that execute the kernel (\ie{} thread grid).
It sets the size of the thread grid and threadblock using built-in three-dimensional variables \textcode{gridDim} and \textcode{blockDim}, respectively. 
Programs access each dimension in these variables through the x, y, and z axes. 
For example, setting \textcode{gridDim.x} as $4$ and \textcode{blockDim.x} as $64$ launches a kernel with a one-dimensional thread grid having four threadblocks, each with $64$ threads ($256$ threads in total).

In the kernel, each thread has access to built-in variables \textcode{threadIdx} and \textcode{blockIdx} with x, y, and z axes. 
The values of these identifiers are bound by the dimensions of the thread grid set by the host. 
For example, \textcode{threadIdx.x} is always less than \textcode{blockDim.x}.
For example, if \textcode{blockDim.x} is set to $64$, the maximum value of \textcode{threadIdx.x} is $63$. 

The kernel follows the SIMT (Single-Instruction-Multiple-Thread) programming model, where different threads execute the same instruction, but operate on distinct data items. 
The data item a thread accesses typically depends on a combination of its identifier variables, the dimension of the kernel's thread grid (\eg{} \textcode{blockDim}), and/or constants. 

\myparagraph{Types of memory}
Unlike CPU, \cuda{} exposes many types of memory with different characteristics to the programmer. 

\begin{mybullet}
\item \textbf{Global memory:}
Data structures of this type are allocated in the GPU's onboard memory (HBM/GDDR) and are accessible to all threads of a GPU kernel. 
Global memory is typically allocated on the host code (\eg{} using \textcode{cudaMalloc}). 
The sizes of the allocations often depend on the problem size (\eg{} matrix dimensions). 
Allocations are freed by calling \textcode{cudaFree}. 
\item \textbf{Shared memory:}
Data structures of this type are allocated on the scratchpad of each SM and are shared among threads of a threadblock (thus, the name). 
These data structures are declared with the `\textcode{\_\_shared\_\_}' specifier. 
There are two methods to allocate shared memory. 
\mycircled{1}~\underline{Static shared memory:} If the size of the data structures in shared memory required for a threadblock is known to the programmer at compile time (\ie{} statically), then they are allocated in the kernel code. 
\mycircled{2}~\underline{Dynamic shared memory:} If the size is not known statically, the kernel code declares a \textit{single} buffer in shared memory with `\textcode{extern}' specifier, but its size is allowed to remain unspecified. 
The size of this buffer is passed as a kernel launch parameter at runtime. 
The programmer then assigns this allocation to a data structure or logically partitions it to assign it to multiple data structures in the kernel code. 
%
\item \textbf{Local memory:}
These data structures are private to the thread that allocates them. 
For large data structures, such as arrays, the compiler allocates them to local memory. 
The size of these structures is known at compile time. 
\end{mybullet}

\myparagraph{Memory safety bugs in GPU programs}
Like C, C++, \cuda{} does not check the bounds and the liveness of memory allocations for performance. 
This makes them vulnerable to memory bugs~\cite{sec-analysis,sok-san, google-safety}. 
There are two prominent types of memory safety bugs that can occur in \cuda{} programs:
\begin{mybullet}
\item \textbf{Out-of-bound (OOB)~\cite{softbound,asan,clarmor}:}
Every memory allocation (global/shared/local) is a contiguous range of virtual addresses, \ie{} has a base address and a size. 
An OOB occurs when a program accesses an address beyond the bounds of the intended allocation (\ie{} beyond `base address + size'). 
OOBs can occur in global, shared, and local memory. 
\item \textbf{Use-after-free (UAF)~\cite{cets,cucatch,letmein}:}
This type of bug occurs when a program uses a pointer (\eg{} dereference) to a data structure, beyond its \textit{liveness}. 
For example, data structures in memory are live between the corresponding memory allocation and free calls. 
A UAF occurs when a thread accesses (\textit{use}) an allocation after it was freed. 
\end{mybullet}

In this work, we focus on OOBs, although our tool also detects UAFs in \cuda{} programs for the sake of completeness. 

\section{Limitations of prior works \& New bug type}
\label{sec:motivation}

Several recent works~\cite{computesanitizer,gmod,clarmor,cucatch,gpushield,letmein,gpuarmor} have focused on improving the memory safety of GPU programs by detecting out-of-bound memory access bugs (OOBs). 
Existing techniques rely on dynamic analysis-based \textit{runtime} approaches.
Software-only tools use instrumentation~\cite {cucatch,computesanitizer,clarmor,gmod}, while others rely on new hardware and/or a combination~\cite{letmein,gpuarmor,gpushield} to trace memory allocations and memory accesses to detect OOBs at runtime.
We noticed that prior works are fundamentally ineffective in detecting elusive OOBs that may not manifest regularly. 
Furthermore, we identify a new type of OOB bug in \cuda{} programs that existing techniques are agnostic to. 
We demonstrate these limitations using real-world \cuda{} programs~\cite{hecbench,indigo} and motivate the need for a new approach to detect OOBs. 

\mysubsection{Ineffective against \textit{input-dependent} OOBs}
\label{sec:motivation:input}

We discover the presence of several \emph{input-dependent} OOBs in open-source \cuda{} programs and libraries.
This class of OOBs manifests in an execution of a program \textit{only under specific inputs}, making them elusive and thus challenging to detect during testing. 
Even programs with OOBs may not manifest them during execution \textit{unless} they are executed with specific inputs, making the OOBs input-dependent. 
These elusive OOBs critically threaten the reliability of GPU programs as they are more likely to escape testing. 
However, they can intermittently and unpredictably cause major disruptions to software in deployment. 

Unfortunately, \textit{no existing technique is designed} to detect input-dependent OOBs. 
All prior proposals \cite{cucatch,clarmor,gmod,computesanitizer,gpuarmor,gpushield,letmein} are runtime techniques.
An OOB \textit{must manifest during an execution} for them to detect it.
Consequently, the coverage of the test inputs and \textit{not} the strength of their techniques determines their effectiveness.  

\begin{figure}[t]
\centering
\begin{code}
/* kernel code */
__global__ void advCubature(..., double* cubD, ... ) {
  id = threadIdx.y * 16 + threadIdx.x;
  s_cubD[j][i] = cubD[id];                      // @\textcolor{red}{OOB in cubD}@
  ...
}
/* host code */
void main() {
  cubN = __input();  Nelements = __input();
  cubNp = (cubN + 1) * (cubN + 1) * (cubN + 1);
  h_cubD = cudaMalloc(cubNp * Nelements * 3);
  ...
  advCubature<<<Nelements, dim3(16, 16)>>>(..., h_cubD, ... );
}
\end{code}
\vspace{-.5em}
\caption{Input-dependent OOB in fluid simulation kernel.} 
\label{fig:mot:dyn-flaw-adv}
\vspace{-1em}
\end{figure}

We show that input-dependent OOBs exist in widely used open-source \cuda{} programs and libraries with two examples. 

\myparagraph{Example 1}
\autoref{fig:mot:dyn-flaw-adv} shows a simplified snippet of the fluid advection simulation program drawn from Oak Ridge National Laboratory's HeCBench~\cite{hecbench}. 
The top section shows the kernel code, and the bottom part shows the host (CPU) code. 
We highlight one input-dependent OOB in the code snippet, while the program itself has $19$ input-dependent OOBs. 

In the kernel, each thread reads from \textcode{cubD} in the GPU's global memory and stores the value in the GPU's shared memory (line $4$). 
It indexes into \textcode{cubD} using index variable \textcode{id}. 
An OOB access occurs in the kernel code when the size of the data structure \textcode{cubD} is \textit{smaller} than the possible values that \textcode{id} can assume (line $4$).

Notice that the host code sets the size for \textcode{cubD} based on user inputs (lines $9$-$11$), \ie{} \textit{input-dependent}, and passes it as kernel launch parameter \textcode{h\_cubD} (line $13$). 
The value of the offset variable \textcode{id} used to access \textcode{cubD} is calculated in line $3$ using the thread identifiers, which, in turn, depend on the dimension of the threadblock.
The kernel (\textcode{advCubature}) is launched with a fixed threadblock dimension with a total of $256$ threads ($16\times16$, line $13$).
Therefore, the calculation in line $3$ suggests that the maximum value of \textcode{id} is $255$.
If the user-provided inputs (\textcode{cubN} and \textcode{Nelements}) are such that the computed size of \textcode{cubD} array (line $9$-$11$) is less than $255$, an OOB access occurs in line $4$ -- an \textit{input-dependent OOB}.

For example, if the user sets the values of \textcode{Nelements} and \textcode{cubN} both to $1$, the  \textcode{cubD} will have a size of $24$ while the index \textcode{id} can be up to $255$.
Such input-dependent OOBs are results of the programmer's \textit{implicit} assumptions about the sizes of data structures that may not hold true under some input.
We discovered several such OOBs across many open-source CUDA programs, \eg{} \textcode{axHelm}~\cite{hecbench}.



\begin{figure}[t]
\centering
\begin{code}
__global__ void push(..., int* nlist, int* data1,
                     int* data2, int numv) {
  int i = threadIdx.x + blockIdx.x * blockDim.x;
  if (i < numv) {
    ...
    for (int j = ... /* iterate over all neighbors */) {
      int neighbor = nlist[j];
      atomicMin(&data1[neighbor], data2[i]);  // @\textcolor{red}{OOB in data1}@
    }
  }
}
\end{code}
\vspace{-.5em}
\caption{Input-dependent OOB in graph processing kernel.}
\label{fig:mot:dyn-flaw-graph}
\vspace{-1em}
\end{figure}

\myparagraph{Example 2} 
\autoref{fig:mot:dyn-flaw-graph} shows the code snippet of the \textcode{pushNode} kernel from the Indigo benchmark suite for graph processing~\cite{indigo}. 
For each vertex in a given \textit{input} graph, the kernel computes the minimum value of data associated with its neighbors.  
On line $8$, \textcode{data1} is indexed using the value in \textcode{neighbor}. 
The value of \textcode{neighbor} is determined by reading from another array \textcode{nlist} (line $7$). 
The \textcode{nlist} array is allocated and populated from a user-provided \textit{input} graph (not shown). 

An input-dependent OOB can occur if the value of the variable \textcode{neighbor} falls outside the size of the array \textcode{data1} (line $8$).
The programmer implicitly assumed that the value of \textcode{neighbor} will be within bounds and makes no attempt in the code (host or device) to verify this. 
Thus, the application can generate OOB accesses upon certain input graphs (\eg{} malformed).
We found similar patterns in other programs such as \textcode{pathCompression}~\cite{indigo} and \textcode{populateWorklist}~\cite{indigo}. 


\mysubsection{Agnostic to logical partitioning of allocations}
\label{sec:motivation:fine}

It is imperative to utilize the GPU's shared memory (scratchpad) for good performance. 
It is typical for programmers to allocate a single contiguous chunk of shared memory and then logically partition it into semantically different data structures. 
Such a program construct is often imposed by the constraints of the \cuda{} programming language. 
If the size of shared memory is input-dependent, \ie{} unknown at compilation time (statically), \cuda{} allows programmers to pass the size as a kernel launch parameter (called dynamic shared memory~\cite{cudaShm}).
However, a kernel can accept only \textit{one} such parameter.
Thus, a programmer can allocate \textit{one} chunk of dynamic shared memory per threadblock. 
If the program semantics require multiple data structures in shared memory, programmers must logically partition the allocation into different data structures using pointer arithmetic in the kernel code. 

We observe the possibility of a \textit{new} class of OOBs if GPU memory accesses erroneously transcend the \textit{logical} boundaries of different data structures carved out of a single memory allocation. 
We name such OOBs \textit{intra-allocation OOB}. 

\begin{figure}[t]
\centering
\begin{code}
// kernel code
__global__ void 
sosfilt(..., int *zi, int sections, int width, ...) {
  extern __shared__ int smem[];
  *s_out = smem;                           // First partition 
  *s_zi = &s_out[sections];                // Second partition
  *s_sos = &s_out[sections * width];       // Third partition
  tx = threadIdx.x;
  for ( int i = 0; i < width; i++ )
    s_zi[tx * width + i] = zi[...];
}
// host code
void main() {
  sosfilt<<<..., shm_size>>>(...);
}
\end{code}
\vspace{-.5em}
\caption{Logical partitioning of shared memory (lines $4$-$7$).}
\label{fig:mot:shared_buffer}
\end{figure}


\myparagraph{Example} 
\autoref{fig:mot:shared_buffer} shows the simplified snippet of \textcode{sosfilt} program~\cite{hecbench}. 
It implements second-order-section (SOS) filtering of digital signals, often used in signal processing. 
The top shows the kernel code, and the bottom shows the host code. 
Here, a shared memory buffer (called \textcode{smem}) is declared as \textcode{extern} in line $4$. 
Note that the size of the buffer is not specified.
The program sets its size at runtime through a kernel launch parameter \textcode{shm\_size} (line $14$).

The programmer logically splits the buffer \textcode{smem} into three distinct buffers, \textcode{s\_out}, \textcode{s\_zi} and \textcode{s\_sos} (line $5$-$7$).   
The first partition (\textcode{s\_out}) starts at the beginning of the allocation. 
The second partition (\textcode{s\_zi}) starts at an offset of the parameter `\textcode{sections}.' 
The third partition (\textcode{s\_sos}) starts at an offset of `\textcode{sections} * \textcode{width}.' 
These buffers, although carved out of a single allocation, serve semantically different purposes in the algorithm. 
An OOB is possible when an instruction that was supposed to access one of these buffers, but overflows and accesses other buffers \textit{within} the same allocated region. 
For example, in line $10$, if the index into \textcode{s\_zi} goes past its boundary, it will overwrite the contents of \textcode{s\_sos}.

Existing techniques are agnostic to the logical partitioning of the allocations (buffers). 
Thus, they are fundamentally incapable of detecting such intra-allocation OOBs.
This is because they track only memory allocations and memory accesses, but do not analyze pointer arithmetic used to semantically create multiple buffers from a single memory allocation.

\begin{figure}[t]
\centering
\begin{code}
// kernel code
template <int rd>               // rd available at compile time
__global__ void KalmanFilter(..., double *RQR, double *T, ...) {
  l_RQR[rd], l_T[rd], l_P[rd];  // multiple local memory buffers
  for (i = 0; i < rd; i++) {
    l_RQR[i] = RQR[i];
  }
  ...
}
\end{code}
\vspace{-.5em}
\caption{Presence of multiple buffers in local memory (line $4$).}
\label{fig:mot:local_buffer}
\vspace{-1em}
\end{figure}

\myparagraph{OOBs in local memory}
We also observed possibilities of OOBs in thread-local memory, due to the semantic partitioning of a single memory allocation. 
\autoref{fig:mot:local_buffer} shows a simplified snippet of Kalman filter~\cite{hecbench}, used to estimate a system's state in robotics. 
The program declares multiple buffers in local memory (line $4$). 
An OOB access to \textcode{l\_RQR} on line $6$, which may erroneously write to \textcode{l\_T}, will not be reported by prior works. 
This is because \cuda{} compilation frameworks generate binaries with a single allocation for the entire local memory needed and then partition it into different data structures.
Prior works are agnostic to such partitioning of local memory allocation as they analyze binaries~\cite{cucatch,computesanitizer}.


\mysubsection{Runtime overheads \& hardware modifications}
\label{sec:mot:over-hard}

Existing techniques rely on dynamic analysis that tracks memory allocations and GPU memory access at runtime. 
Consequently, they add significant runtime performance overhead, \eg{} up to $3.2\times$ for \cucatch{}~\cite{cucatch}.
Additionally, software-only tools incur large memory overhead to maintain metadata, thereby limiting the memory available for the programs.
For example, \cucatch{} incurs a memory overhead of $2\times$. 

To limit the overheads, researchers have proposed hardware modifications such as hardware buffers to cache metadata~\cite{gpuarmor} and/or hardware for bounds checking~\cite{letmein}. 
These require extensive modifications to the GPU that do not currently exist and may not necessarily be adopted in the future.

\begin{table}[t]
  \centering
  \scriptsize
  \caption{Qualitative comparison of existing works for detecting OOBs in GPU programs with \tool{} (this work).}
  \label{tab:mot:existing}
  \setlength{\tabcolsep}{5pt}
  \begin{tabular}{l|c|c|c|c}
    \textbf{Tool Name}                         & \textbf{Miss Input}      & \textbf{Ignore intra-}             & \textbf{Runtime}  & \textbf{Modified} \\
                                               & \textbf{dependent OOB}   & \textbf{allocation OOB}            & \textbf{overhead} & \textbf{hardware} \\ \hline
    Comp. San~\cite{computesanitizer}          & Yes & Yes & High & No  \\
    \cucatch{}~\cite{cucatch}                  & Yes & Yes & High & No  \\
    GMOD~\cite{gmod}                           & Yes & Yes & High & No  \\
    clARMOR~\cite{clarmor}                     & Yes & Yes & High & No  \\ \hline
    GPUShield~\cite{gpushield}                 & Yes & Yes & Low &  Yes \\
    LMI~\cite{letmein}                         & Yes & Yes & Low &  Yes \\
    GPUArmor~\cite{gpuarmor}                   & Yes & Yes & Low &  Yes \\ \hline
    \textbf{\tool{}}                           & No  & No  & -   &  No  \\ \hline
  \end{tabular}
\end{table}

\mysubsection{Summary: Comparison with prior works}
\autoref{tab:mot:existing} summarizes the qualitative comparison between prior work and our proposal, \tool{}. 
The top part lists software-only tools, while the ones in the second half require hardware modifications.
Finally, the last row lists the software tool we created, called \tool{}. 

From the first column, we note that \textit{all} prior works are fundamentally ineffective against input-dependent OOBs. 
This limitation arises because all prior works require the OOBs to manifest to be detected. 
As mentioned in the second column, \textit{all} prior works fail to report intra-allocation OOBs. 
While the software-only tools suffer from high overheads ($>$$2\times$) incurred due to instrumentation, they do not require hardware modification. 
Tools with low runtime overheads ($<$$1.1\times$) require significant modifications to key components of GPU hardware. 
In the next section, we discuss how \tool{} overcomes these limitations and the key insights behind it. 

\section{Goals and Design Principles}
\label{sec:design}

We set out with three key goals: 
\mycircled{1} Detect \textit{input-dependent} OOBs \textit{without} requiring them to manifest. 
\mycircled{2} Detect \textit{intra-allocation} OOBs. 
\mycircled{3} Avoid runtime overheads and the need for modified hardware. 

To achieve goals \mycircled{1} and \mycircled{3}, we avoid dynamic (runtime) analysis, unlike prior works.
This is a fundamental requirement, as the outcome of dynamic analysis applies to a specific execution of the program under a particular input. 
Thus, it is incapable of proactively detecting input-dependent OOBs that are yet to manifest. 
Instead, we propose a first-of-its-kind static (compile-time) analysis technique.
It analyzes both host and kernel code to identify semantic relations between host variables that determine memory allocation sizes and kernel variables that determine the offsets of memory accesses to those allocations.
It then uses a SAT solver~\cite{sat-wiki} to determine if the offsets can exceed the size of allocations. 

To achieve goal \mycircled{2}, we propose to analyze an intermediate representation (IR), unlike prior works that analyze program binaries. 
The IR retains program-level semantics about the logical partitioning of allocations into different data structures, unlike binaries. 
We created a compile-time tool, named \tool{}, that embodies these techniques. 
Next, we describe its key design principles and how these help \tool{} mitigate the limitations described in \autoref{sec:motivation}. 

\mysubsection{Semantic relations for detecting input-dependent OOBs}
\label{sec:design:method}

The key challenge is to determine if an OOB can occur in a given program under \textit{any possible input allowed (legal) by the program}.\footnote{The legal input is a set of inputs that a program can accept without being restricted by constructs such as asserts in the program.}
Importantly, it must do so at the time of compiling the program, when the inputs that the program may encounter during its execution are not available.

We make a key observation to overcome this challenge -- \textit{even if the program's inputs are unknown during compilation, the size of the allocations and offsets (indices) of memory accesses into those allocations must be semantically related for OOBs to not exist}.
We observe that such semantic relations occur naturally in bug-free GPU-accelerated programs. 
For example, the size of allocated buffers in the GPU memory (\eg{} using \textcode{cudaMalloc)} is typically derived from the input and/or the problem size (\eg{} matrix dimensions). 
Furthermore, programmers would naturally set the dimension of the thread grid based on the same input/problem size, thereby relating the allocation size to the thread-grid dimension. 

An input-dependent OOB arises when the programmer fails to programmatically establish a relation that binds the range of indices within the size of the allocation it accesses. 

We then observe that the offset values of a memory access into an allocation are almost always derived from thread and threadblock identifiers of the individual threads (\eg{} \textcode{threadIdx}, \textcode{blockIdx}). 
This stems fundamentally from the GPU's SIMT programming model, where each thread is expected to operate on different data items.
Consequently, a thread's unique identity determines the data item(s) it operates on. 
Importantly, the range of values that these identifiers can assume during an execution is bounded by the thread grid dimension set during kernel launch. 
This, in turn, \textit{semantically binds} the offsets of memory accesses to the sizes of the corresponding memory allocation, \emph{irrespective of the program's inputs}.
If a programmer fails to embed such a relation in the code that would restrict the offsets to fall within the allocation sizes, an input-dependent OOB occurs.

Importantly, \emph{semantic relations between the allocation sizes and the offsets are inferable from the source, thanks to the use of common program variables for deriving them}. 
Specifically, the sizes of the allocation and the thread-grid dimensions are typically derived from common variables in the \emph{host code}. 
The offsets into the memory allocations are calculated from the thread, and/or block identifiers in the \emph{kernel code}. 
Finally, the kernel launch parameters relate these host variables, thread grid dimension, and kernel variables together. 

\begin{figure}[t]
\centering
\begin{code}
// kernel code
__global__ void saxpy(int* array, int devNumElements) {
  index = threadIdx.x + blockDim.x * blockIdx.x;
  array[index] = ...;
}
// host code
void main() {
  multiples = __input();
  blkDim = 512;         // 512 elements per block
  gridDim = multiples;  // number of blocks
  array = cudaMalloc(multiples * blkDim);
  saxpy<<<gridDim, blkDim>>>(array, multiples * blkDim);
}
\end{code}
\caption{Semantic relation among variables determining allocation size and thread grid dimensions (lines $9$-$12$).}
\label{fig:design:host_device}
\end{figure}

We elucidate this by taking the example of a typical code snippet shown in \autoref{fig:design:host_device}.
Here, the actual size of the memory allocation (\textcode{array}) is not known at compilation.
It depends on user input (lines $8$-$9$, host code).
Similarly, the thread grid dimension also depends on the same input.
It is tied to the allocation size through the use of common host variables, \textcode{numElements} and \textcode{blkDim}.
The kernel variable \textcode{index} in the kernel code is the offset of accesses to the \textcode{array}.
The value of the \textcode{index} is calculated by each thread at runtime based on their respective thread and block identifier (line $3$).
The value of these variables is bounded by the grid dimension, which in turn is dictated by the host variables \textcode{numElements} and \textcode{blkDim} in the kernel launch parameters. 
Finally, the \textcode{array} in the kernel code is related to the memory allocated in the host code \textcode{array\_d} through launch parameters (line $12$). 

Driven by these observations, \tool{} analyzes the host and kernel code to trace the variables used in allocating memory (\textcode{cudaMalloc}), setting the thread grid dimensions, and indexing into the allocations. 
It further traces assert statements that can limit the values of these variables. 

The next challenge is to ascertain if offsets of memory accesses \textit{can} go beyond the allocation sizes under \textit{any} legal input.
Towards this, we utilize a SAT (satisfiability) solver~\cite{sat-wiki}. 
We create three kinds of constraints (equalities or inequalities) that must be satisfied for a static instance of memory access to possibly cause an OOB. 
\mycircled{1} The value of the thread and block identifiers must be less than the grid dimensions -- true by \cuda{} semantics.
\mycircled{2} We encode the semantic relations among the program variables inferred by analyzing the IRs in the form of equations.
These constraints must be true for the given program for \textit{every input}.
\mycircled{3} Finally, for a memory access to cause an OOB, its offset must be greater than or equal to the size of the memory allocation it accesses under some input. 
We encode this as an inequality constraint.
The SAT solver then checks if all these constraints are \textit{simultaneously} satisfiable under some input. 
If yes, the given memory access can cause an OOB.
Otherwise, it cannot under any input.

For \textit{every} static instance of memory access instruction in the kernel, \tool{} checks if it may cause an OOB in the aforementioned way.
As this technique does \textit{not} rely on actual input, but only on the relation between the variables and constraints that must always be true, it is capable of detecting all types of OOBs, including the elusive input-dependent ones. 


We illustrate how this approach helps find input-dependent OOBs in \autoref{fig:mot:dyn-flaw-adv}. 
The size of the data structure \textcode{cubD} depends upon inputs captured in host code variables \textcode{cubN} and \textcode{Nelements} (lines $9$-$11$).
The variable, \textcode{Nelements}, also determines the number of threadblocks to be launched. 
However, the threadblock dimension is fixed ($16\times16$) and is \textit{unrelated} to the input and thus, the allocation size. 
Further, \textcode{h\_cubD} in the kernel launch parameter relates to \textcode{cubD} in the kernel code.
The variable \textcode{id} in the kernel code is used for indexing into \textcode{cubD}.
Each thread calculates its value using their identifiers (line $3$).
Thus, the threadblock dimension bounds the range of values that it can assume. 
However, the dimension is fixed -- the programmer failed to relate it to input variables that determine the size of \textcode{cubD}.
As a result, the SAT solver finds that offsets (\textcode{id}) can assume values greater than the size of \textcode{cubD}, \ie{} a possible OOB.

\mysubsection{Inferring logical partitions to find intra-allocation OOBs}
\label{sec:design:partition}

The next challenge is to detect OOBs in the presence of logical partitions, where a memory allocation is divided into distinct data structures. 
We address this in two steps. 
\mycircled{1} Unlike prior works that analyze binaries at runtime, \tool{} analyzes the intermediate representations (IR). 
IRs retain key semantic information, unlike the compiled binary. 
It tracks instructions that perform pointer arithmetic on variables pointing to memory allocation or its derivatives. 
This helps find the logical partitions and their boundaries within each allocation (if any).
\mycircled{2} We then create constraints for the SAT solver that determine if a memory instruction can possibly access outside its \textit{logical partition}, following the same methodology discussed earlier.
The solver then checks the satisfiability of these constraints, thereby finding the possibility of intra-allocation OOBs. 

For example, in the code snippet in \autoref{fig:mot:shared_buffer}, the IR corresponding to lines $6$-$7$ performs arithmetic on the variables holding the base pointer of a memory allocation (\textcode{s\_out}) and its derivative (\textcode{s\_zi}, \textcode{s\_sos}). 
The analysis determines the bounds of the logical partitions derived from variables such as \textcode{sections}, \textcode{width}, and pointer variables (\textcode{smem}, \textcode{s\_out}, and \textcode{s\_zi}). 
We then create inequalities (constraints) for offsets of memory accesses (\eg{} \textcode{tx*width + i}, line $10$) to go past the bounds of the logical partition it accesses (\eg{} \textcode{s\_zi}). 
The solver checks the feasibility of these additional constraints, along with the rest of the constraints discussed in the previous section, to identify possibilities of OOBs. 

Finally, we observe that program binaries contain only \textit{one} allocation for all data structures allocated in local memory. 
For example, in \autoref{fig:mot:local_buffer}, the kernel allocates multiple  local memory buffers in line $4$ (\textcode{l\_RQR}, \textcode{l\_T}, \textcode{l\_P}). 
However, the compiled binary allocates a single contiguous chunk of local memory for each thread.
Consequently, existing dynamic analysis techniques are oblivious to individual data structures allocated in local memory. 
In contrast, \tool{} analyzes the IR that retains individual allocations in local memory, allowing for the tracking of individual allocations. 
\tool{} then uses the SAT solver and the constraints derived from the common variables, as usual, to ascertain the existence of OOBs for accesses to data structures in the local memory.

\mysubsection{Detecting use-after-free errors}
\label{sec:design:overheads}

While we focus on OOBs, \tool{} also detects temporal safety bugs, such as use-after-free (UAF) errors, for the sake of completeness.
Specifically, it detects UAF errors in programs that can manifest under any possible input.  

\tool{} analyzes the IR of host and kernel code to detect memory allocations and de-allocations. 
It maintains a set of \textit{live} (valid) memory allocations at any point in the program.
At the start of the program, this set is empty. 
Upon encountering an allocation, it is added to the set.
Upon encountering a de-allocation, the corresponding entry is removed from the set. 
On encountering a \textit{use}, we check if the corresponding allocation (data structure) is in the live set. 
We define \textit{use} as a memory access in the kernel, or the allocation passed as a kernel launch argument in the host code. 
If an allocation is not in the live set at the time of a \textit{use}, a UAF is reported.

\section{Implementing \tool{}}
\label{sec:impl}

\begin{figure}[t!]
    \centering
    \includegraphics[width=\linewidth]{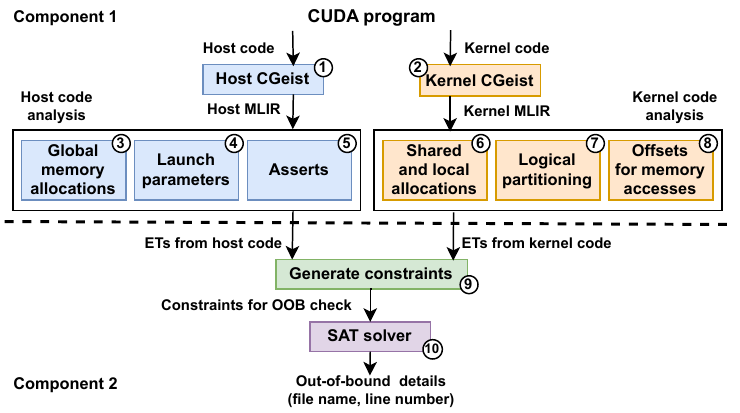}
    \vspace{-1.5em}
    \caption{High-level components of \tool{}.}
    \label{fig:impl:main}
    \vspace{-.5em}
\end{figure}

\autoref{fig:impl:main} shows the high-level working of \tool{} that detects out-of-bounds (OOB) in \cuda{} programs. 
The input \cuda{} program is processed by \tool{}'s two key components. 
First, it compiles the program, consisting of host and kernel code, using CGeist~\cite{polygeist} to MLIR intermediate representation (IR)~\cite{mlir} (\mycircled{1} and \mycircled{2} in the figure). 
MLIR provides an IR structure that retains significant semantic information from the source code and a mature infrastructure for analyzing the IR. 
\tool{} extracts semantic relations between different program variables by analyzing the host code's MLIR (\mycircled{3}-\mycircled{5}) and kernel code's MLIR (\mycircled{6}-\mycircled{8}) (\autoref{sec:impl:analysis}). 
Next, \tool{} creates informed constraints from the analysis (\mycircled{9}) and uses Google OR-Tools' SAT solver~\cite{ortools} to test the satisfiability of the constraints (\autoref{sec:impl:constraints}) in \mycircled{10}. 
Additionally, \tool{} analyzes the host and kernel code to detect UAFs~\cite{cucatch} (\autoref{sec:impl:uaf}). 
Finally, it reports the details of each instruction in the kernel code (\ie{} filename, line number) that may participate in memory safety bugs (if any). 

\subsection{Extracting semantic relation among program variables}
\label{sec:impl:analysis}

\tool{} analyzes the IR of the host code to extract the variables that determine the size of global memory allocations and the thread grid dimensions (\mycircled{3} and \mycircled{4} of \textcode{Component} in \autoref{fig:impl:main}).
It analyzes the IR for the kernel code to capture static shared/local memory allocations (\mycircled{6}), logical partitioning of shared memory allocations (\mycircled{7}), and offsets of the memory accesses (\mycircled{8}). 
\tool{} captures this information by constructing an \textit{expression trees} (ETs)~\cite{expressionTrees} for the relevant program variables and expressions. 

\myparagraph{Expression trees}
We first detail how expression trees (ETs) are created, as they lie at the heart of extracting the semantic relations from the IR. 
The ETs capture the \emph{equation} that calculates the values of program variables during execution in the form of a binary tree.  
The leaf nodes of the tree represent a constant or an input variable whose value is unknown at the time of compilation (\textcode{Unknown}). 
Internal nodes of the tree represent a binary operator.
The root represents how the value of a variable is calculated at runtime.

\begin{figure}[t!]
    \centering
    \includegraphics[width=.6\linewidth]{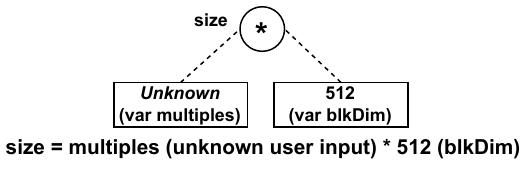}
    \caption{Expression tree (ET) for \textcode{size} in \autoref{fig:design:host_device}, line $11$.}
    \label{fig:impl:et}
\end{figure}

\autoref{fig:impl:et} depicts the ET for the size of memory allocation (\textcode{cudaMalloc}) in the code snippet of \autoref{fig:design:host_device}. 
The size of the allocation is determined by multiplying the variables \textcode{blkDim} and \textcode{multiple}. 
While the former is a constant ($512$), the latter is an input variable and thus unknown (\textcode{Unknown}). 
The root node captures the multiplication operation between the two to arrive at the allocation size.

\begin{figure}
\begin{code}
ET createET(Statement st) {
  // recursively create ET for statements and its constituents
  if (isConstant(st)) {                 // constant ET
    return ET { st.ConstantValue };
  } else if (isUserInput(st)) {         // user input ET
    return ET { Unknown };
  } else if (isBinaryOperation(st)) {   // binary ET
    leftEt = createET(st.lhs);
    rightET = createET(st.rhs);
    return ET { leftET st.operator rightET }; 
  } else if (/* other possibilities */) { ... }
}
\end{code}
\caption{Simplified pseudo-code to create expression trees (ET).}
\label{fig:impl:express:create}
\vspace{-1em}
\end{figure}

\autoref{fig:impl:express:create} shows the simplified pseudo-code (\textcode{createET}) to create ET for an MLIR statement that defines a variable. 
It first checks if the statement is a constant value or a user-provided input (lines $3$-$6$). 
It returns the constant value for the former and \textcode{Unknown} for the latter. 
If the statement is a binary operation, then it recursively calls itself for the left and right operands. 
It creates a node with the operator in the statement with the results of the recursive calls as children (lines $7$-$10$). 

Importantly, the relationships among the statements (\eg{} the size of a memory allocation and the thread grid dimension) that arise from the use of common program variables are naturally captured across ETs. 
Each variable definition in MLIR has a \textit{unique} identifier, called a \textcode{Value} object. 
In creating ETs, \tool{} uses the \textcode{Value} object to represent a variable across different ETs uniquely. 
Thus, any use of common program variables is captured in the ETs of related statements. 

\begin{figure}[t!]
\begin{subfigure}{.5\textwidth}
\centering
\begin{code}
map<Addr,ET> allocs;     // tracking global memory allocations
// each kernel's grid dimension, arguments, dynamic shared mem
list<<Kernel, list<ET>, list<ET>, ET>> kernelInfo;
list<ET> asserts;                         // assert conditions
if (isMemoryAlloc(/* Statement */st)) {   // e.g., cudaMalloc
  allocs[st.Ret] = createET(st.size);
} else if (isKernelLaunch(st)) {
  for (dim in st.GridDim())               // kernel dimensions
    gridDims.push(createET(dim));
  for (arg in st.KernelArgs()) {          // kernel arguments
      if (isPointer(arg))                 // pointer type
        args.push(allocs[arg]);
      else
        args.push(createET(arg));         // e.g., scalar type
  }
  if (st.hasDynShm())
    dynShm = createET(st.DynShm())        // dyn. shared memory
  kernInfo = { st.KernelName, gridDims, args, dynShm };
  // track all kernel information
  kernelInfo.add(kernInfo);
} else if (isAssert(st)) {
  asserts.push(createET(st.condition));   // assert condition
} else if (/* other host code Statements */) { ... }
\end{code}
\caption{Simplified pseudo-code for analyzing host code's MLIR.}
\vspace{.5em}  
\label{fig:impl:express:host}
\end{subfigure}
\hfill
\begin{subfigure}{.5\textwidth}
\centering
\begin{code}
map<Addr,ET> kAllocs;       // local/static shared allocations
list<<Addr,ET>> partitions;               // partitions offset
list<tuple<Addr,ET>> memInstrs;           // mem instructions 
if (isStaticAlloc(/*Statement*/ st)) {    // kernel allocations
  kAllocs[st.Ret] = createET(st.Op);
} else if (isDynShAlloc(st)) {            // base of dyn shm
  partitions.push({ st.Ret, createET(null) });
} else if (isCreatePartition(st)) {       // logical partitions
  partitions.push({ st.Ret, createET(st.Op) });
} else if (isMemInstr(st)) {              // memory access
  memInstrs.push({ st.ds, createET(st.offset) });
} else if (/* other operations */) { ... }
// create ET for size of each partition
for (curPart: partitions.size() - 1) {
  // difference between next offset and current offset
  kAllocs[curPart.Addr] = { nextPart.ET - curPart.ET };
}
// add base and ET of size for last partition to kAllocs
kAllocs[lastPart.Addr] = { kern.dynShm - lastPart.ET } 
\end{code}
\caption{Simplified pseudo-code for analyzing kernel code's MLIR}
\label{fig:impl:express:kernel}
\end{subfigure}
\caption{Creating ETs from host and kernel code analysis.}
\label{fig:impl:express}
\vspace{-1em}
\end{figure}

\myparagraph{Host code analysis}
\tool{} first generates the MLIR of the host code using CGeist~\cite{polygeist} (\mycircled{1} in \autoref{fig:impl:main}).
It then analyzes the MLIR to extract ETs for the size of global memory allocations, kernel launch parameters (\eg{} thread grid dimension), and that of assert statements (if any).
\autoref{fig:impl:express:host} shows the simplified pseudo-code for our MLIR-based compiler pass, which analyzes the host code. 

\begin{mybullet}
\item \textbf{Global memory allocation (\mycircled{3} in \autoref{fig:impl:main}):} 
Upon encountering a call to \textcode{cudaMalloc}, \tool{} creates the ET of the size argument of the allocation (lines $5$-$6$).   
It stores the ET into a map (\textcode{allocs}) indexed by the MLIR \textcode{Value} object of the allocation variable (unique identifier).
\item \textbf{Launch parameters (\mycircled{4}):} 
Upon encountering a kernel launch (\textcode{LaunchOp} in MLIR), it parses the launch parameters, including the thread grid dimension, kernel arguments, and dynamic shared memory (if available). 
For each axis of the grid dimension, it creates the ET of the variable for the size of that dimension and stores it in a list (lines $8$-$9$).
For each kernel argument, it first checks if the argument is a pointer, \ie{} points to a global memory allocation. 
If yes, it looks up the \textcode{allocs} map to find the ET of the allocation's size and adds the ET to the list of arguments (\textcode{args}) (lines $11$-$12$). 
Otherwise, it creates the ET of the argument and adds it to the list (line $14$). 
If the kernel uses dynamic shared memory, it creates the ET for its size (lines $16$-$17$).
Finally, \tool{} adds all the information about the kernel launch, including the kernel's name, to the \textcode{kernelInfo} list (lines $18$-$20$).
\item \textbf{Asserts (\mycircled{5}):} 
Upon encountering an assert statement, \tool{} creates ET of the assert condition and stores it in the \textcode{asserts} list (lines $21$-$22$). 
\end{mybullet}

\myparagraph{Kernel code analysis}
\tool{} generates MLIR for the kernel code (\mycircled{2} in \autoref{fig:impl:main}).
It analyzes the MLIR to track the sizes of shared and local memory allocations (\mycircled{6}), logical partitioning of dynamic shared memory allocations (\mycircled{7}), and offsets of memory access instructions (\mycircled{8}).
\autoref{fig:impl:express:kernel} shows a simplified pseudo-code that analyzes the kernel MLIR.  

\begin{mybullet}
\item \textbf{Shared and local memory allocations (\circled{6} in \autoref{fig:impl:main}):} 
The \textcode{alloca} statement in MLIR is responsible for allocating both shared and local memory. 
The first argument of \textcode{alloca} specifies the size of the allocation, if known at the compilation, as is the case for static shared memory and local memory allocations (\textcode{isStaticAlloc}, line $4$).  
The second argument specifies the type of allocation to differentiate between shared and local memory. 
However, for the purpose of detecting OOBs, this distinction is irrelevant and is ignored. 
\tool{} creates ET for the size argument of the allocation and stores it in a map data structure (\textcode{kAllocs}) indexed by the unique MLIR \textcode{Value} identifier of the allocation (lines $4$-$5$). 

\item \textbf{Logical partitioning (\mycircled{7}):}
The \textcode{alloca} statement is also used in MLIR to allocate dynamic shared memory.
However, the size argument is unspecified (`\textcode{?}') in the MLIR. 
Upon detecting such allocation, \tool{} adds the MLIR \textcode{Value} that uniquely identifies the allocation and a \textcode{NULL} ($0$) ET into a \textcode{partitions} list (lines $6$-$7$). 

The MLIR statement \textcode{SubIndexOp} performs the pointer arithmetic to logically partition an allocation.
\autoref{fig:impl:mlir} shows an example snippet in MLIR's IR for logical partitioning of dynamic shared memory, compiled for the kernel code in \autoref{fig:mot:shared_buffer}.  
The \textcode{SubIndexOp} statement takes two arguments --- pointer to the base allocation and the offset from the base where the partition begins (line $4$, \autoref{fig:impl:mlir}). 
For each \textcode{SubIndexOp}, \tool{} creates the ET of the offset argument and adds it to a list (\textcode{partitions}), along with its unique \textcode{Value} identifier (line $8$-$9$ in \autoref{fig:impl:express:kernel}). 


After finishing the pass, processing the kernel code's MLIR, \tool{} iterates over the list of partitions (lines $14$-$19$) to create ETs for their \textit{size}. 
The size of a partition is the difference between the offset of the next partition and its offset.
The starting offset is zero (\textcode{NULL}). 
The tool captures this by creating a new ET having a root node with a subtraction operator (`-') and the ET of the next entry in the \textcode{partitions} list as the root's left child and with the current entry's ET as the right child (line $16$). 
This new ET captures the program variables and structure of calculation that determines the size of a partition.
Notice that the size of the \textit{last} partition depends on the size of the dynamically shared memory.
Recall that \tool{} captures the ET for this size from the host code analysis. 
It uses this to create the ET for the size of the last partition (line $19$).

\item \textbf{Offsets for memory accesses (\mycircled{8}):}
\tool{} scans the kernel code's MLIR for static memory instructions (\ie{} loads, stores, and atomics).
It tracks two operands of each memory instruction -- the variable pointing to the data structure it accesses (\textcode{st.ds}) and the offset (index) into the data structure (\textcode{st.offset}).
It then creates the ET of the offset. 
This ET, along with the variable pointing to the data structure, is inserted into the \textcode{memInstrs} list (lines $10$-$11$). 
The variable, \textcode{st.ds}, could point to a data structure on the shared memory (or one of its partitions) or local memory allocated by the \textcode{alloca} statement in the kernel's MLIR.
It could also point to a data structure in global memory. 
For the latter, the \textcode{st.ds} will have the positional argument in the kernel launch parameter. 
This helps uniquely identify host code allocated (\textcode{cudaMalloc}-ed) data structures that are passed to the kernel through kernel launch parameters. 
For the purpose of detecting OOBs, the place of residence of a data structure is irrelevant. 
\end{mybullet}

\begin{figure}[t]
\centering
\begin{assemcode}
...
...
\end{assemcode}
\vspace{-.5em}
\caption{Simplified MLIR snippet for program in \autoref{fig:mot:shared_buffer}.}
\label{fig:impl:mlir}
\vspace{-.5em}
\end{figure}

\mysubsection{Detecting OOBs using a SAT solver}
\label{sec:impl:constraints}

Finally, \tool{} uses Google OR-Tools' SAT solver~\cite{ortools} to ascertain if a memory access instruction can cause an OOB. 
For every memory instruction in the kernel's MLIR, it provides a set of constraints that must all be satisfied simultaneously for it to cause an OOB (\mycircled{9} in \autoref{fig:impl:main}). 
The solver explores the search space of possible values of variables to either declare that the set of constraints cannot be satisfied together (no OOB) or provide values that would satisfy the constraints (OOB) (\mycircled{10}). 
A GPU kernel is free of OOB if none of its memory accesses can cause an OOB.  

\begin{figure}[t!]
\begin{subfigure}{.49\linewidth}
    \centering
    \includegraphics[width=\linewidth]{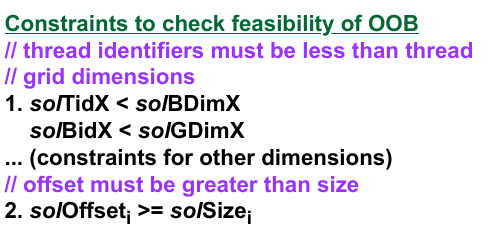}
\end{subfigure}
\begin{subfigure}{.49\linewidth}
    \centering
    \includegraphics[width=\linewidth]{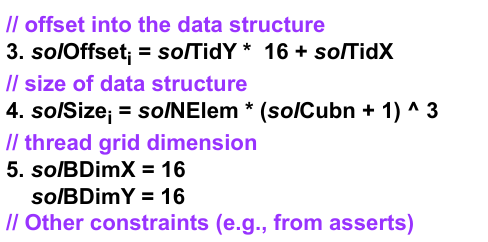}
\end{subfigure}
\vspace{-.5em}
\caption{Constraints to check for OOBs in instruction `i.'}
\label{fig:impl:constraint}
\vspace{-1em}
\end{figure}

\tool{} expresses constraints as inequalities or equalities among \textit{solver variables} and/or constants.
Solver variables represent program variables in the solver's search space, \ie{} they can take any legal values during the solver's check for satisfiability. 
\tool{} creates \textit{three} kinds of constraints. 
The first kind encodes the constraints that the values of thread and threadblock identifiers (positive integers) \textit{must be less than} the threadblock and grid dimensions along each of the three axes (x, y, z).
These constraints are always true as necessitated by \cuda{} semantics.
The item (1) of \autoref{fig:impl:constraint} shows an example of such constraints. 
The prefix `\textit{sol}' indicates a solver variable.
Terms \textcode{TidX} represent \textcode{threadIdx} in the `x' dimension, while \textcode{BidX} represents the \textcode{blockIdx}. 
Similarly, \textcode{BDimX} represents the number of threads in a threadblock in the `x' dimension while \textcode{GDimX} represents the number of threadblocks in the grid. 

Next, \tool{} creates constraints to test the feasibility of an OOB for each memory access instruction in the kernel (second type of constraints). 
It iterates through the \textcode{memInstrs} list, which contains all memory instructions. 
For each entry, \tool{} creates the constraint that the offset of the access must be greater than or equal to the size of the data structure it accesses. 
This is a necessary condition for the given memory instruction to cause an OOB. 
The item (2) of \autoref{fig:impl:constraint} depicts one such constraint.
The subscript `\textit{i}' enumerates the solver variable for the offset (\textcode{solOffset}) and that for the size (\textcode{solSize}) of the $i^{th}$ memory instruction in \textcode{memInstrs}. 

Finally, \tool{} creates the third kind of constraints that capture how the program calculates relevant variables and relations among them.
First, as it iterates over the list of memory instructions (\textcode{memInstrs}), it creates an equality constraint capturing how the offset of the instruction is computed. 
Specifically, the ET for the offset from the entry in \textcode{memInstrs} is expressed as an equation of solver variables. 
Item (3) of \autoref{fig:impl:constraint} depicts such a constraint. 
It shows the expression for the offset calculation in line $3$ of the code snippet in \autoref{fig:mot:dyn-flaw-adv}.

Next, \tool{} adds an equality constraint to capture the calculation of the size of the data structure that the given instruction accesses. 
For that, \tool{} must first locate the ET for the allocation size of that data structure. 
It looks up the \textcode{kAllocs} dictionary with the identity of the data structure from the instruction's entry in \textcode{memInstrs}.
The ET of the size of allocation will be available in the \textcode{kAllocs} if the data structure is allocated by the kernel onto the shared memory or local memory. 
If it is allocated in the global memory by the host code, then the lookup will miss. 
In that case, the entry of the \textcode{memInstrs} would contain the positional argument of the kernel's launch parameter. 
It is then used to access the corresponding entry in the \textcode{args} list of the \textcode{kernelInfo} that contains the kernel arguments. 
The entry in this list has the ET for the size of the data structure. 
\tool{} then expresses the ET of the size of the data structure (on shared, local, or global memory) as an equation of solver variables. 
The item (4) of \autoref{fig:impl:constraint} depicts such a constraint representing line $11$ of the code snippet in \autoref{fig:mot:dyn-flaw-adv}.

\tool{} then iterates over the list of ETs in \textcode{gridDim} that captures the calculation of thread grid dimensions in the host code.
These ETs are then expressed as equations that must be satisfied.
The items (5) in \autoref{fig:impl:constraint} show these constraints for the example in \autoref{fig:mot:dyn-flaw-adv}.
Finally, it iterates over the list of ETs for assert conditions (\textcode{asserts}) and expresses each as an equation of solver variables that the solver must satisfy. 

The solver searches the space of all possible assignments of values to its variables that can simultaneously satisfy all the constraints for each of the memory instructions. 
If it finds such an assignment, the given memory access can cause an OOB.
Otherwise, it cannot. 
If none of the instructions can cause an OOB, then the program is free of any spatial memory safety error. 
In the example of \autoref{fig:impl:constraint}, the SAT solver would correctly find out that it is possible to satisfy all constraints simultaneously, thus detecting a possibility of an OOB without needing it to manifest, unlike prior works.

\mysubsection{Detecting use-after-free bugs}
\label{sec:impl:uaf}

While our contribution is for detecting OOBs in GPU programs, \tool{} also detects UAFs~\cite{cucatch,letmein,gpuarmor}. 
Given the MLIR, \tool{} checks if the \textit{use} of an allocation (data structure) is \textit{live} (valid). 
An allocation is \textit{used} in memory access (kernel code) or as arguments in a kernel launch (host code). 
A data structure is live at a given point in execution if it is allocated \textit{and} not freed before its \textit{use}. 

\tool{} iterates over statements in the kernel MLIR to check if memory instructions access live data structures. 
It maintains a set of live data structures (allocations) which is initially empty. 
Upon encountering an allocation, it adds the allocation to the set. 
On encountering a de-allocation, it removes that allocation from the set. 
If allocations or de-allocations appear under conditional statements, \tool{} builds the live set for the \textcode{if} path and the \textcode{else} path before taking their intersection to update the live set. 
This ensures that \tool{} can catch UAFs irrespective of the \textcode{if}/\textcode{else} branch taken during execution. 
For each memory instruction, \tool{} checks if the corresponding allocation is in the live set. 
If not, it reports a UAF. 

\tool{} also iterates over the host code's MLIR and maintains a live set of allocations. 
On encountering a kernel launch, \tool{} checks if all pointer arguments are in the live set. 
Otherwise, it reports a UAF.
\section{Evaluation}
\label{sec:eval}

We evaluate \tool{} to assess its ability to detect the two classes of elusive OOBs -- \mycircled{1} the input-dependent OOBs, and \mycircled{2} the intra-allocation OOBs. 
We also evaluate its propensity to report false positives, a common vulnerability of any static analysis tool. 
Finally, we discuss the overheads of \tool{}.

We evaluate using a broad range of workloads from popular open source repositories such as Kaldi~\cite{kaldi}, llm.c~\cite{llmC}, \cuda{} samples~\cite{CUDASamples}, and benchmark suites such as ScoR~\cite{scor}, Rodinia~\cite{rodinia}, HeCBench~\cite{hecbench}, and Indigo~\cite{indigo}. 

\myparagraph{Comparison with prior works} 
\lmi{}~\cite{letmein}, \gpuarmor{}~\cite{gpuarmor}, GPUShield~\cite{gpushield} require hardware modification and thus, simulators for evaluation. 
We cannot quantitatively compare them with our software tool. 
Further, their artifacts are unavailable. 

Unfortunately, functional artifacts of prior software tools, cuCatch~\cite{cucatch} and GMOD~\cite{gmod}, are not available. 
Even the binaries of these tools were not made available. 
The clARMOR~\cite{clarmor} is a software tool applicable only to OpenCL programs, while we focus on \cuda{}.
Thus, two are incompatible. 

The only publicly available tool for detecting memory safety bugs in \cuda{} programs is NVIDIA's \csan{} (\csanshort{})~\cite{computesanitizer}.
Importantly, though, the bug-finding abilities of all prior works are similar to \csanshort{}'s -- they differ only in terms of performance.
For example, \textit{all} prior works, including \csanshort{}, are runtime techniques and are \textit{not} designed to detect input-dependent and intra-allocation OOBs. 
In short, in terms of bug finding, \csanshort{} fairly represents all prior works. 

We evaluated \tool{} on a system with an Intel Core i5-14400F CPU, $64$GB of DRAM, and an NVIDIA $4080$ GPU.
It runs on Ubuntu v$24.10$, and CUDA v$12.1$.
We used the latest commit of CGeist~\cite{polygeist} as of August $2025$ and LLVM release $18$ for MLIR infrastructure.

\subsection{Analyzing \tool{}'s ability to detect OOBs}
\label{sec:eval:effective}

\begin{table}
  \centering
  \scriptsize
  \caption{OOB detection result of tools. Each cell mentions the total number of real OOBs (R), introduced OOBs (I), number of false positives (P), and number of false negatives (N) reported by tools in the format `R/I (P/N).'}
  \label{tab:eval:detection}
  \begin{tabular}{l|c|c}
    \textbf{Application}                     & \textbf{\csan{}} & \textbf{\tool{}}     \\ \hline
    fluidAdvection~\cite{hecbench}           &  0/0 (0/19)        &  19/0 (0/0)        \\
    axHelm~\cite{hecbench}                   &  0/0 (0/2)         &  2/0 (0/0)         \\
    luDecomposition~\cite{rodinia}           &  0/0 (0/2)         &  2/0 (0/0)         \\ 
    pathCompression~\cite{indigo}            &  0/0 (0/3)         &  3/0 (0/0)         \\
    populateWorklist~\cite{indigo}           &  0/0 (0/2)         &  2/0 (0/0)         \\
    pushNode~\cite{indigo}                   &  0/0 (0/2)         &  2/0 (0/0)         \\ \hline 
    sosfil~\cite{hecbench}                   &  0/0 (0/2)         &  0/2 (0/0)         \\ 
    matmul~\cite{cudaPtx}                    &  0/0 (0/4)         &  0/4 (0/0)         \\ 
    softmax~\cite{llmC}                      &  0/0 (0/3)         &  0/3 (0/0)         \\ 
    kalmanFilter~\cite{hecbench}             &  0/0 (0/6)         &  0/6 (0/0)         \\ \hline
    globalOOb~\cite{cucatch}                 &  0/1 (0/0)         &  0/1 (0/0)         \\
    localOOb~\cite{cucatch}                  &  0/1 (0/0)         &  0/1 (0/0)         \\
    staticSharedOOb~\cite{cucatch}           &  0/1 (0/0)         &  0/1 (0/0)         \\
    dynamicSharedOOb~\cite{cucatch}          &  0/1 (0/0)         &  0/1 (0/0)         \\ \hline
    CopyUpperToLower~\cite{kaldi}            &  0/0 (0/0)         &  0/0 (0/0)         \\
    CopyLowerToUpper~\cite{kaldi}            &  0/0 (0/0)         &  0/0 (0/0)         \\
    CopyFromMat~\cite{kaldi}                 &  0/0 (0/0)         &  0/0 (0/0)         \\
    NeedlemanWunsch~\cite{rodinia}           &  0/0 (0/0)         &  0/0 (0/0)         \\
    convolutionSeparable~\cite{CUDASamples}  &  0/0 (0/0)         &  0/0 (0/0)         \\
    rule110~\cite{scor}                      &  0/0 (0/0)         &  0/0 (0/0)         \\ \hline
    \textbf{Summary}                         &  0/4 (0/45)        &  30/19 (0/0)       \\ \hline
  \end{tabular}
\end{table}

\autoref{tab:eval:detection} compares OOBs reported by \csanshort{} and \tool{} over a diverse set of programs. 
The programs are separated into four sections. 
The top section lists programs that have input-dependent OOBs. 
These are programs or kernels from popular libraries that have OOB but were \textit{not} reported to date, \ie{} real OOBs (R) that remained undetected. 
The next section lists the programs that logically partition dynamic shared memory or have multiple buffers in local memory. 
They have the program structure that could cause intra-allocation OOB, but they did not have real OOBs. 
We slightly extended them to introduce OOBs (I) without perturbing their program structure. 

The third section lists micro-benchmarks with OOBs on global, local, and shared memory that the existing tools can detect without fail. 
We use them to demonstrate that both \csanshort{} and \tool{} can detect such OOBs. 
Finally, the last section lists programs that do \textit{not} have OOBs. 
These programs evaluate \tool{}'s ability to avoid false positives, \ie{} false alarms when an OOB does \textit{not} exist.
Together, these programs test the robustness and accuracy of \tool{}s' ability to detect OOBs, \ie{}  ability to report OOBs \textit{without} false negatives (missing actual OOB) and false positives. 

Each cell of \autoref{tab:eval:detection} reports the total number of real OOBs in unmodified programs (R), introduced OOBs (I), the number of false positives (P), and false negatives (N), reported by a given tool for a given program in the format `R/I (P/N).'
A tool should strive to limit the number of false positives and false negatives. 
The last row summarizes the OOBs reported by each tool, along with the number of false positives and false negatives (lower is better). 

From the first section of \autoref{tab:eval:detection}, we notice that \csanshort{} is unable to detect any of the input-dependent OOBs and thus, suffers from false negatives, as expected.
Like all prior works, it relies on runtime analysis and requires the OOBs to manifest in order for it to detect them.
However, many input-dependent OOBs are \textit{elusive} -- they manifest only under specific inputs. 
For example, in the \textcode{luDecomposition} program~\cite{rodinia}, an OOB manifests only if the input matrix size is \textit{not} a multiple of $16$. 
In contrast, \tool{} relies on semantic relations between the allocation sizes, thread grid dimensions, and offset calculations for memory access to detect all \textit{30 previously unreported} input-dependent OOBs across $six$ programs.


All prior techniques, including \csanshort{}, are agnostic to OOBs that occur in the presence of logical partitions of dynamic shared memory or across multiple local buffers (\autoref{sec:motivation:fine}). 
They are oblivious to pointer arithmetic operations that create multiple logical data structures from a single memory allocation. 
Several programs, \eg{}  \textcode{sosfil}, \textcode{softmax}, \textcode{matmul}, logically partition a single allocation on shared memory into multiple data structures. 
The application, such as \textcode{kalmanFilter}, \textcode{fluidAdvection}, allocates multiple local buffers. 
While these applications are susceptible to intra-allocation OOBs, they do not have actual OOBs. 
Thus, we induced OOBs in them (`I' in \autoref{tab:eval:detection}). 
For example, in \autoref{fig:mot:shared_buffer}, we incremented the index into \textcode{s\_zi} by $1$ (line $10$). 
As the index is still within the bounds of the allocation, \csanshort{} cannot detect them. 
However, \tool{} analyzes IR that preserves the semantics of logical partitioning and multiple local memory buffers. 
As a result, it reports intra-allocation OOBs. 

Next, input-\textit{independent} OOBs in the set of microbenchmarks are detected equally well by both \csanshort{} and \tool{}. 
We expect other existing tools to detect them as well. 

Finally, we evaluated \csanshort{} and \tool{} against programs that do \textit{not} have OOBs to quantify their propensity to report false positives (last section in \autoref{tab:eval:detection}). 
Since \csanshort{} detects an OOB only when it manifests, it cannot generate false positives. 
Thanks to the accurate inference of semantic relations among program variables, \tool{} also does not report any false positives. 
From the summary row, we note that \csanshort{} has many false negatives ($45$ in total), whereas \tool{} catches all OOBs, including input-dependent and intra-allocation OOBs. 

We also evaluated \tool{} to detect UAFs using programs from prior works~\cite{cucatch,letmein} (not listed in the table). 
These programs include UAFs, such as accessing a freed allocation in the kernel and passing a freed pointer during kernel launch. 
Both \csanshort{} and \tool{} detect all UAFs in these programs. 


\subsection{Analysis time of \tool{}}
\label{sec:eval:time}

\tool{} adds \emph{no} runtime performance or memory overheads, unlike prior works. 
However, being a static analysis tool, it lengthens the compilation time. 
Importantly, though, it is a one-time cost.   
The compilation time ranged from eight milliseconds to at most $32$ seconds (longest for \textcode{copyFromMat} program). 
The SAT solver contributed the highest share of the compilation time. 
However, this is a one-time cost incurred during the debugging phase and does not impact the program's runtime.  
In contrast, \csanshort{} is a binary instrumentation-based tool with runtime overheads ranging from $1.1\times$ and $4\times$ as it traces every memory access and allocations at runtime. 
\section{Related Work}
\label{sec:related}

\myparagraph{GPU memory bugs detection}
NVIDIA's \csan{}~\cite{computesanitizer}, GMOD~\cite{gmod,gmodx}, clARMOR~\cite{clarmor,clarmor-cgo}, and \cucatch{}~\cite{cucatch} perform instrumentation to detect OOBs. 
They incur high runtime performance (up to $4\times$) and/or memory overheads (up to $2\times$). 
GPUShield~\cite{gpushield}, GPUArmor~\cite{gpuarmor}, and LMI~\cite{letmein} perform hardware-assisted base and bounds checks to detect OOBs. 
IMT~\cite{imt} and LAK~\cite{lak} maintain tagged memory allocations and pointers in hardware to check if a pointer accesses a memory location with a different tag. 
All these techniques rely on dynamic analysis and are ineffective against input-dependent OOBs. 
They are not designed to detect intra-allocation OOBs, either. 
\tool{} overcomes these flaws through static analysis and the use of SAT solvers. 

\myparagraph{CPU memory bugs detection}
Valgrind~\cite{valgrind}, Softbound~\cite{softbound}, Address Sanitizer~\cite{asan}, and Range Sanitizer~\cite{RSan} perform instrumentation to detect memory bugs at runtime. 
Their reliance on runtime analysis makes them ineffective against input-dependent OOBs. 
Model checkers have been used for finding memory bugs in CPU programs~\cite{cbmc,esbmc}. 
They model the execution of each thread in the program. 
However, this approach is not scalable for GPU programs executing thousands of concurrent threads. 
%
SPARC~\cite{sparc}, ARM MTE~\cite{arm} detect memory bugs through hardware-assisted allocation and pointer tagging, while REST~\cite{rest}, Califorms~\cite{califorms} use trip-wires. 
CHERI~\cite{cheri,cherivoke,cherix86}, Intel MPX~\cite{mpx}, No-FAT~\cite{nofat}, and AOS~\cite{aos} maintain the base and bounds of allocations in special data structures in hardware to detect OOBs. 
These techniques rely on dynamic analysis and thus are susceptible to missing input-dependent OOBs. 
However, thanks to static analysis, \tool{} detects input-dependent OOBs. 
\section{Conclusion}
\label{sec:conclusion}

Existing tools to detect memory bugs in GPU-accelerated programs are ineffective against input-dependent OOBs and cannot detect intra-allocation OOBs. 
We observe that the semantic relation among variables that determine the size of memory allocations and the index into these allocations can be used to detect the presence (or absence) of OOBs. 
Further, analyzing kernel code to track logical partitioning of memory allocations provides the necessary information to detect intra-allocation OOBs. 
We build \tool{}, which leverages these observations to detect elusive memory bugs in GPU kernels.

\bstctlcite{IEEEexample:BSTcontrol}
\bibliographystyle{IEEEtranS}
\bibliography{sample-base}

\end{document}